\documentclass[sigconf]{acmart}

\AtBeginDocument{%
  }

\usepackage{graphicx} 
\usepackage{hyperref}
\usepackage{enumitem}
\usepackage{pifont}
\usepackage{ulem}

\usepackage{subcaption}
\usepackage{amsthm}
\usepackage{booktabs}    
\usepackage[table]{xcolor} 

\usepackage{longtable}
\usepackage{array}

\newcommand{\para}[1]{\vspace{2pt}\noindent\textbf{#1.~}}

\renewcommand{\subsubsection}[1]{%
  \refstepcounter{subsubsection}%
  \vspace{0.3em}\noindent\textbf{\thesubsubsection\quad #1}\par
  \vspace{0.3em}
}

\newcommand\designer{UI designer{}}
\newcommand\generator{\textsc{P2dGen{}}}
\newcommand{\ignore}[1]{}

\renewcommand\footnotetextcopyrightpermission[1]{} 

\title{Designing UIs with GenAI: Understanding the Use, Benefits, and Limitations of Prompt-to-Design Generators}

\author{%
  Jingzhou Ye,
  Yao Li,
  Xueqiang Wang
}
\affiliation{%
  \institution{University of Central Florida}
  \city{Orlando}
  \state{Florida}
  \country{USA}
}
\email{%
  {jingzhou.ye, yao.li, xueqiang.wang}@ucf.edu
}

\begin{document}

\begin{abstract}

The integration of generative AI (GenAI) into user interface (UI) design has led to prompt-to-design generators (\generator{}s), tools that enable designers to create UI prototypes from textual prompts efficiently. 
Despite their growing adoption, little is known about how designers use these tools in everyday practice, their benefits, and limitations. 
We conducted semi-structured interviews with 17 UI designers to investigate \generator{}s use, focusing on workflow integration, attitudes toward generator-assisted design, and the organizational climate.
Findings show that designers primarily use \generator{}s as ideation accelerators rather than end-to-end solutions, and they adopt a variety of individual and collaborative prompting strategies.
Designers expressed complex attitudes, noting gains in efficiency and inspiration while also highlighting concerns about loss of personal creativity and design quality. 
We also observed emerging governance practices and varying attitudes within organizations toward \generator{}s. 
These insights illuminate how \generator{}s reshape UI design workflows and contribute to understanding human-AI collaboration in UI design.

\end{abstract}



\maketitle


\section{Introduction}
\label{sec:intro}





The integration of GenAI tools is reshaping established user interface (UI) design practices.
A recent survey indicated that 89\% of designers use some form of GenAI tools in routine UI design tasks~\cite{weavely2024ai}.
One of the most commonly used types of tools is the \textit{prompt-to-design generators} (\generator{}s), such as Uizard~\cite{uizard2024}, Vercel's V0~\cite{vercel2025v0} and Figma Make~\cite{figma2024make}. These tools promise to accelerate designers move from concept to prototype. As a result, UI design is no longer solely a manual process driven by sketches, wireframes, and iterative refinement, but an emerging form of human, AI collaboration in which generative systems actively participate in design decision-making.

Despite the rapid adoption of \generator{}s, there remains limited designer-centric understanding of how these generators are actually used in real-world UI design workflows, as well as their benefits and limitations.
Prior work~\cite{huang2021creating, yuan2025towards, yuan2025designrepair} has largely focused on the technical capabilities of generative systems or their output quality, while paying less attention to how designers integrate them into existing practices, how these tools affect the UI design process, and how designers work with \generator{}s. 
Without such understanding, it remains unclear how these tools are actually being used in and how they would potentially change existing UI design workflows.
In addition, UI designers’ attitudes toward adopting generative UI tools, like their perceptions and expectations, as well as the organizational climates influencing the further adoption of \generator{}s, such as the norms, guidelines, and governance mechanisms surrounding their use in companies, remain underexplored.
Taken together, these factors make it difficult to assess the long-term value of UI design with \generator{}s and to inform the responsible and improved design of future tools.
%

To address these gaps, this study investigates the impact of \generator{}s on design workflows from individual, social, and organizational perspectives, through semi-structured interviews with 17 UI designers who use these tools.
Specifically, we aim to answer the following research questions: 

\textbf{RQ1: }\textit{How do designers currently use generators in UI design workflows?}

\textbf{RQ2: }\textit{What are users’ attitudes toward adopting generators in UI design?}

\textbf{RQ3: }\textit{How does organizational climate influence the adoption and governance of generators?}

Our findings reveal insights into several previously less known aspects of how \generator{}s are used, perceived, and may change today’s UI design practices.
First, to interact with \generator{}s, UI designers leverage individual strategies for creating prompts, such as those driven by prompt templates, UI design contexts, and user needs, as well as strategies for using prompts, including repeated prompting and multimodal prompting.
They also adopt collaborative prompting strategies, such as prompt co-creation with UI designers, project managers, and software engineers, as well as designer–AI collaboration for creating prompts.
This variety of prompting strategies potentially suggests a lack of agreed-upon best practices and indicates opportunities to optimize interactions with \generator{}s.
Second, UI designers expressed ambivalent and sometimes conflicting attitudes toward adopting \generator{}s in UI design. 
While participants valued \generator{}s for improving efficiency, accelerating early-stage exploration, and providing sources of inspiration, they at the same time voiced concerns about diminished personal creativity, erosion of design craftsmanship, and uneven or superficial design quality in generated outputs. 
Third, at the organizational level, we identify divergent organizational stances and emerging governance practices surrounding \generator{} adoption. 
Some organizations actively encouraged experimentation with \generator{}s and framed them as strategic assets for innovation, while others adopted cautious or restrictive approaches due to concerns about intellectual property, data security, brand consistency, and accountability.
Across these contexts, designers navigated evolving policies, informal norms, and unclear boundaries regarding when and how \generator{}s should be used, highlighting how organizational climate plays a critical role in shaping both the uptake and the regulation of \generator{}s in design work.

Our work aims to provide a grounded understanding of how \generator{}s are reshaping design practice, how designers experience and interpret these changes, and how organizational factors mediate the integration of AI into design work. These insights contribute to ongoing discussions in HCI and design research about the future of creative work in the era of GenAI, and offer implications for tool designers, design leaders, and organizations seeking to responsibly adopt \generator{}s.  The study involving UI designers was reviewed and approved by the IRB of our institution. 
\section{Related Work}
\label{sec:related}


\subsection{GenAI Tools Used in UI Design}
\para{Prompt-to-UI generation tools (\generator{})}
Prior research has long pursued the goal of generating UI design artifacts (e.g., interface layouts, low-fidelity mockups, or full prototypes) from natural language descriptions of intended UIs.
For example, earlier studies~\cite{huang2021creating} explored how deep neural networks, such as Generative Adversarial Networks (GANs), can generate low-fidelity UI mockups from high-level textual descriptions.
The widespread availability of large language models (LLMs) has further enabled the development of such prompt-to-UI generation tools (in short we call \generator{}s).
Examples include studies~\cite{feng2023designing, yuan2025towards, yuan2025designrepair} that support the generation of mid-fidelity wireframes, or improved human-AI interaction or UI repair capabilities, and Text2App~\cite{hasan2021text2app} that generates full applications with multiple UI flows.
Similar tools have also been adopted by industry practitioners, including Figma Make~\cite{figma2024make}, Uizard~\cite{uizard2024}, etc.
A recent large-scale survey indicates that 89\% of researchers adopt GenAI tools, including \generator{}s, in routine UI design tasks~\cite{weavely2024ai}, highlighting the growing need to understand how these tools concretely affect UI design.
More recent research on UI design with GenAI has begun to address challenges in the controllability and interpretability of purely prompt-driven generation. 
Misty~\cite{lu2025misty} supports interactive conceptual blending, letting designers merge selected aspects (e.g., layout, color) of a reference UI into a work-in-progress design with a transparent semantic diff. 
SQUIRE~\cite{leung2025squire} replaces open-ended prompting with explicitly scoped, point-and-click component tree editing, to help determine what will and will not be modified in the generated UIs. 
Athena~\cite{beason2026athena} scaffolds multi-screen app generation through shared intermediate representations: a storyboard, data model, and per-screen GUI skeletons. 
These systems suggest that structured, explicitly scoped interactions could help address the attribution concerns described by participants in this study (Section~\ref{sec:discussion}), though these studies evaluate researcher-built systems rather than commercial \generator{}s in everyday use as we do.

\para{Use of LLMs in other UI/UX design tasks}
Beyond directly generating UI mockup prototypes, prior studies have also explored the use of LLMs in other aspects of design work.
For example, Duan et al.~\cite{duan2024generating} demonstrated that LLMs can evaluate UI mockups and identify usability issues in a manner comparable to traditional heuristic evaluation methods.
PromptInfuser~\cite{petridis2024promptinfuser} showed that LLMs can provide useful UI design suggestions, which improve iteration speed and support designers’ decision-making when using design tools.
Similarly, Park et al.~\cite{park2025leveraging} leveraged multimodal LLMs to support inspirational UI search, helping designers retrieve relevant reference designs from natural language and visual queries rather than directly generating a finished UI.
Another line of research has examined the use of LLMs to assist software developers who write UI code.
For instance, DesignCoder~\cite{chen2026designcoder} introduces hierarchy-aware UI code generation techniques to improve fidelity in multi-level UI structures.
UICoder~\cite{wu2024uicoder} investigates how human feedback can be used to fine-tune LLMs to generate front-end UI code that is better aligned with design intent and platform conventions.
Recent work has also explored using LLMs to generate interfaces directly for end-user interaction, rather than to support designers or developers. 
Chen et al.~\cite{chen2026generative} propose a paradigm in which LLMs dynamically synthesize task-specific interactive interfaces, such as data visualizations and interactive tools, in place of static text, using structured representations (e.g., interaction flow graphs and finite-state machines) with iterative, reward-driven refinement.
Overall, these efforts demonstrate the broad applicability of LLMs across UI/UX-related tasks, such as generating feedback to improve UI designs and producing UI code from existing designs to support software developers.
While these efforts contribute to improved UI/UX, they are not directly concerned with the design process itself (but rather with feedback generation or code synthesis) and are therefore not the focus of this study.

\subsection{UI Design Workflows Shaped by GenAI}
\para{UI design workflow in the pre-GenAI era}
Before the rise of generative AI, UI design followed an iterative workflow~\cite{uiuxdesign_process_Jaye, uiuxdesign_process_Maria} that moved from understanding user needs to delivering and refining functional interfaces. Designers typically began with user and market research to uncover motivations, context, and business goals. Based on these insights, they then defined user personas, information architecture, and user journeys to shape the conceptual direction of the product.
Following this, designers produced wireframes, visual layouts, and interactive prototypes. Usability testing played a critical role in revealing issues and informing revisions to enhance the user experience. Once the design reached maturity, it was handed off to developers for implementation and further iteration.
A variety of UI/UX tools supported this workflow, such as Miro~\cite{miro2025} for mapping user flows and Figma~\cite{figma2025} for interface design and prototyping, among others.
Despite the availability of these tools, the success of the design still largely depended on designers’ expertise and creative judgment to transform abstract requirements into well-crafted user interfaces. Their ability to interpret user insights, balance aesthetic and functional considerations, and make nuanced trade-offs throughout the design process remained irreplaceable. Designers needed to navigate ambiguous problem spaces, explore multiple design alternatives, and apply design principles such as consistency, accessibility, and visual hierarchy. In other words, tools served primarily as enablers, while the core creative and decision-making responsibilities continued to rest on designers themselves.

\begin{figure*}[h]
  \centering
  \includegraphics[width=0.65\linewidth]{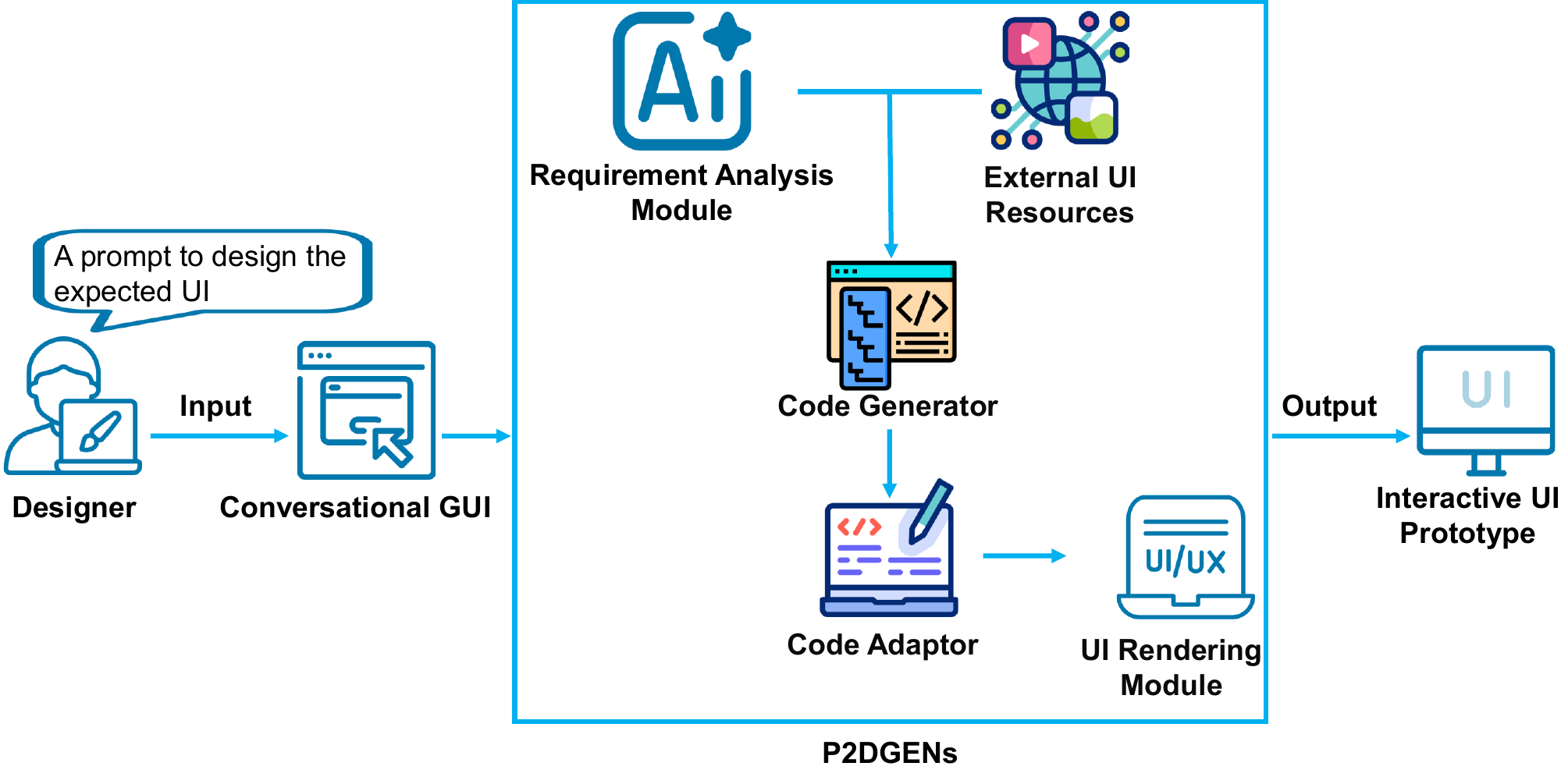}
  \caption{A brief overview of a \generator{} and its components}
  \label{fig:pipeline}
\end{figure*}

\para{UI design with \generator{}s}
With \generator{}s, \designer{}s can simply provide prompts describing the intended UI requirements and directly obtain interactive prototypes, usually in the form of front-end code and rendered interfaces. Although different \generator{}s may vary in their underlying mechanisms, they typically include several essential components, as shown in Figure~\ref{fig:pipeline}.
At the interaction level, \generator{}s offer a \textit{designer conversational GUI} through which designers input prompts that articulate the desired UI. The prompts are processed by a \textit{design requirement analysis module}, which extracts key semantic elements, such as the UI’s purpose, functional needs, and stylistic preferences.
After this analysis, \generator{}s leverage LLMs commonly to synthesize front-end code capable of constructing a prototype. This step may involve selecting and assembling external UI resources (either drawn from a built-in repository or retrieved from the web), including component libraries, pre-designed templates, and image assets. When necessary, \generator{}s dynamically adapt these resources to the given prompt, for instance, adjusting styling to match the extracted aesthetic specifications.
Once the front-end code is produced, a \textit{UI rendering module} is invoked to visualize the prototype for \designer{}s to evaluate. If revisions are needed, \designer{}s can iteratively refine the design by submitting additional prompts through the \textit{designer conversational GUI}, continuing the prompt-driven improvement cycle.

\subsection{Understanding GenAI Use in UI/UX Design}
The integration of AI into UI/UX design workflows has been studied from multiple angles, revealing both significant opportunities and persistent tensions.
In 2022, Lu et al.~\cite{lu2022bridging} conducted a need-finding study with eight UX professionals and identified a fundamental research-practice gap: while AI-based tools have advanced rapidly in academia (such as for tasks like layout generation, design search, and UI generation), practitioners rarely adopt them (at the time of the paper writing).
Their findings highlighted four areas where AI assistance could be genuinely useful: design inspiration search, design alternative exploration, design system customization, and automated guideline violation checking.
Critically, they found that most GenAI tools operate in overly simplified scenarios and produce outputs too generic to fit practitioners' real-world branding and contextual constraints, while also failing to support higher-order "design thinking" activities such as user interviews and brainstorming.
Shortly after the above study, in 2023, Shi et al.~\cite{shi2023understanding} conducted a systematic literature review of 93 papers and organized the designer–AI relationship into three themes: GenAI assisting designers through discovering, visualizing, creating, and testing; designers assisting AI through training and regulating; and five dimensions characterizing their collaboration, including scope, access, agency, flexibility, and visibility.
Their review further underscored the need for more explainable, ethical, and adaptable AI, and echoed concerns about the limited adoption of research prototypes in practice.

More recently, with the emergence of LLMs and their enhanced capabilities, in 2025, Khan et al.~\cite{khan2025beyond} examined AI's role specifically in the divergent thinking stages of UI/UX design through interviews with 19 professional designers.
They found that designers value GenAI tools that kick-start creativity, support research synthesis, generate design alternatives, and enable rapid prototyping, but insist on retaining creative control throughout. Notably, designers preferred AI as a collaborative assistant that frees them from repetitive tasks to focus on creative exploration, rather than as an autonomous generator.
They also identified a blurring boundary between low-fidelity and high-fidelity prototyping as GenAI tools rapidly produce realistic outputs.
Kamnerddee et al.~\cite{kamnerddee2024ai} conducted a direct empirical comparison of human-created and AI-generated UI prototypes to complement the qualitative accounts,  for a fitness mobile application, using A/B testing and the System Usability Scale.
Their results found no statistically significant difference in usability scores between the two prototype types, yet participants consistently preferred the human-created design for intuitiveness and seamlessness, suggesting that AI-generated outputs, while efficient, still fall short in nuanced, human-centered design understanding.
Beyond examining the divergent thinking stages and output quality, Takaffoli et al.~\cite{takaffoli2024generative} interviewed UX practitioners to understand how generative AI is used across UX design, as well as how company- and team-level policies shape its adoption.
Besides the level of individual designers, other work has examined how GenAI reshapes collaboration at the team level.
Subramonyam et al.~\cite{subramonyam2025prototyping} observed collaborative prototyping sessions with GenAI tools across product teams and found that prompt engineering and interface design become intertwined: UX designers, PMs, and engineers jointly negotiate how prompts should constrain the interface, and technical decisions once centralized to engineers become accessible to non-technical roles.

However, these studies have focused on general-purpose GenAI tools, such as ChatGPT~\cite{chatgpt}, DALL-E3~\cite{dalle3}, and Midjourney~\cite{midjourney}, including a recent study by Naqvi et al.~\cite{naqvi2025catalyst} comparing designers' perceptions of such tools across career stages, which are not specifically designed for UI design and primarily function as conversational or ideation assistants rather than systems that explicitly generate UI artifacts.
In contrast, \generator{}s directly produce concrete UI designs from prompts, introducing different interaction paradigms, constraints, and design responsibilities for designers. For instance, designers may need to more precisely encode design intent, contextual constraints, and user-centered considerations into prompts that drive automated UI generation.
Consequently, the challenges, strategies, and implications of adopting \generator{}s differ fundamentally from those associated with general-purpose GenAI tools.
Despite this, little is known about how designers integrate \generator{}s into their existing workflows, what prompting strategies they develop individually and collaboratively, how they navigate trade-offs between efficiency and creative autonomy, and how organizations govern their use.
Perhaps closest to our work is Chen et al.'s Generative UI (GenUI) Study~\cite{chen2025genui}, a week-long diary study in which 37 professionals across four roles used a single GenUI tool for individual mini-projects. 
Our work differs in scope: rather than assigning one tool for a fixed duration, we recruit designers with prior, self-directed experience across multiple commercial \generator{}s, which allows us to capture prompting strategies and organizational governance practices as they emerged organically in everyday practice.
Our work addresses this gap by offering a grounded account of \generator{}s as a distinct and increasingly prominent form of human–AI collaboration in UI design.
\section{Methodology}
\label{sec:methodology}

We conducted a semi-structured interview study with 17 participants to address the three research questions about designers’ usage and attitude of the \generator{}s. 

\subsection{Recruitment of Participants}
We recruited participants in the United States who were 18 years or older and had prior experience applying \generator{}s in UI design. This requirement ensured that all participants contributed insights grounded in actual practice with AI-assisted design rather than speculative or unfamiliar usage.
To access a wide pool of designers, we shared the recruitment announcement across several online UI design communities, including the Facebook group “UI \& UX Designers”~\cite{facebook_uiux} and the Reddit forum “UI\_Design”~\cite{reddit_uiux}. We further expanded our reach through participant referrals, enabling us to include designers outside these online groups.
During the qualitative analysis of interview data, we monitored for thematic saturation—defined as the point at which no new codes emerged. Recruitment concluded once saturation was reached, resulting in a total of 17 participants (P1–P17). Among them, 16 participants were recruited through social media and one through referral.
All potential participants first completed a screening survey delivered via Qualtrics~\cite{qualtrics_free_account}. The survey opened with an informed consent form, followed by eligibility questions (e.g., age and previous use of \generator{}s in UI design). It then collected demographic information, including gender, education level, and years of UI design experience, to ensure diversity in the sample. Participants who met the criteria were asked to provide an email address for scheduling the study session.
Each study session lasted approximately 60 minutes and was conducted remotely via Zoom. Participants received a \$40 Amazon gift card as compensation. The study protocol received approval from our institution’s IRB and followed all IRB guidelines. Participant demographic details are provided in Table~\ref{tab:participant_info} in the Appendix.

\subsection{Interview Process}
Each consenting UI designer was invited to participate in a scheduled Zoom session, following the workflow described below. At the start of the session, we greeted the participant, introduced the research team, and outlined the overall study flow. We also explicitly stated that there were no right or wrong answers, and that participants’ responses would not be evaluated during the session. We then conducted a set of warm-up questions to gather contextual information about their UI design background, including their team or company size, their role within the organization, the design tools they routinely relied on (both traditional tools and GenAI options), and how frequently \generator{}s were incorporated into their existing workflows. Participants were also asked to elaborate on their use of \generator{}s more broadly, such as usage patterns, motivations, perceived challenges, and the level of contribution they felt AI made to their design process.
%
%
Following the introduction, we probed deeper into participants’ experiences and perspectives on using \generator{}s in UI design. 
%
%
Specifically, we explored the types of prompts participants commonly used, the considerations and strategies they applied when crafting prompts, and the key information they deemed essential to include (e.g., design goals, user context, constraints, and stylistic preferences). We also asked about participants’ workflows with \generator{}s, such as how they achieve desired design and control output quality, as well as the challenges and difficulties they encountered during the workflows.
The interview topics also included questions addressing organizational perspectives on AI adoption, such as workplace policies, governance considerations, and whether enterprise versions of \generator{}s were available or supported.
During the probe, we also welcomed participants to screen share and demonstrate their prompts in \generator{}s. 
After this, we closed each Zoom session by thanking the participant and then distributed compensation after the meeting concluded. 
The full list of interview questions is in Table~\ref{tab:interview_list} in the Appendix.


\subsection{Data Analysis}
We analyzed the interview data using inductive thematic analysis~\cite{braun2006using}. All audio-recorded interviews were first transcribed and anonymized, and the original audio files were deleted once transcription was complete to ensure participant privacy. After preparing the dataset, we thoroughly reviewed all materials, including interview transcripts and screen-sharing recordings, to familiarize ourselves with the full scope of the data.
Each of the three researchers in our group independently generated initial codes, focusing on participants’ understandings, perceptions, practices, challenges, and strategies related to the use of \generator{}s. 
First, the three researchers independently coded the first five transcripts to generate candidate codes, which were subsequently consolidated into an initial codebook comprising 66 codes. The remaining transcripts were then coded using this codebook while allowing new codes to emerge, and data saturation was reached after P14.
These codes represented meaningful labels attached to specific excerpts of the data. Throughout the process, we met repeatedly to compare and reconcile our codes, iteratively returning to the raw dataset to refine our interpretations. Through this collaborative process, we produced a consolidated set of 235 codes.
Building on these codes, we grouped related ones into preliminary sub-themes, ultimately identifying five sub-themes. For each sub-theme, we compiled the relevant data excerpts, examined how the associated codes related to one another, and merged overlapping codes into broader conceptual patterns. 
For example, P5 noted that he will discuss prompts with groups, and P10 mentioned that he will use other GenAI tools to help write prompts. We categorized both under the pattern “Assisted and Collaborative Strategies.” We then observed that this pattern shares similar characteristics with “Individual-Level Strategies,” and therefore grouped both under the subtheme “Prompting Strategies for UI Design.” 
These sub-themes were then further organized into higher-level themes, yielding three overarching themes. For instance, the subthemes “Adoption of \generator{}s in UI Design” and “Assisted and Collaborative Strategies” both describe how participants use \generator{}s, so they were merged into the same theme. Together, these themes, sub-themes, and codes formed a thematic map that illustrated the structure of our findings, consisting of three themes and five sub-themes.
Finally, we reviewed and refined the thematic map to ensure that the themes accurately captured the meanings represented across coded data segments and collectively formed a coherent and interpretable analytic narrative.

\section{Results}
\label{sec:results}

\subsection{Designers’ Usage of \generator{}s  in UI Design Workflows}

\subsubsection{Adoption of \generator{}s in UI Design}

%
In total, participants reported using a wide range of AI-assisted design tools. 
Across all responses, the tools mentioned fell primarily into three categories: \generator{}s (e.g., Figma Make, Uizard, Lovable and Framer AI), visual and color-palette assistants (e.g., Khroma), and general-purpose AI creative tools (including Canva and conversational agents such as ChatGPT and Claude). 
Notably, all participants reported relying on at least one \generator{}.
%
%
%
Thirteen (76.5\%) participants (P1–P9, P14–P17) 
%
%
reported using \generator{}s frequently in their workflows, indicating a common shift from traditional UI designs to \generator{}-assisted designs.
%
%
For example, P14 noted:

\begin{quote}
    \textit{``I use both Figma Make and other AI tools to help with my everyday tasks, which normally involves me trying to be creative in the aspect of designing.''} 
\end{quote}

%

\para{Factors influencing the adoption of \generator{}s}
The first factor is that designers rely on \generator{}s more in new projects than in existing ones.
Several participants (P4, P5, P10) pointed out that the extent to which \generator{}s are used largely depends on the type of project. 
For instance, they mentioned that they rely more heavily on \generator{}s for UI design when working on new projects, but find it difficult to apply these tools to existing projects due to challenges in replicating existing UI designs and making improvements to them.
These findings highlight a potential area for improving \generator{}s -- they perhaps should not only support UI generation from scratch but also enhance their ability to ingest and make targeted changes to existing UIs.
%
%
%
For example, P5 noted:

\begin{quote}
\textit{
``If we bring a project from scratch, we use it [\generator{}s] predominantly for everything. If a project has already a pre-existing design, we actually have to tweak a lot more than we like, because we need to get it to the current state and then use AI to modify it.''
}
\end{quote}
\ignore{
P10 further explained that even when AI is used in mature projects, it is mainly applied at the ideation stage rather than for direct reuse:

\begin{quote}
\textit{``Most of the time it is used more like at the beginning for exploration, for example, to try different design patterns, or to give you a general direction to try. It’s more like an idea or ideation-stage reference, but it cannot be used directly.''}
\end{quote}
}

The second factor is that designers use \generator{}s differently depending on their roles.
For example, designers’ roles shape their priorities when using \generator{}s: UX designers prioritize workflow efficiency due to their focus on user interactions and task completion, while graphic designers emphasize UI aesthetics, reflecting their focus on visual design and style.
Thus, participants (P4, P5, P6, P16) integrate AI in ways that align with what they prioritize in their roles. 
P5 explained this distinction as follows:

\begin{quote}
\textit{``
Different people care about different things. There are people who care about the visuals and they need to get pixel perfect designs. There are also people, especially in the UX flows, they couldn’t care much about whether this has 100 or 200 pixels padding. They care about, are we showing the right words? Does the user get the right instructions?
''}
\end{quote}

This role-based difference is also reflected in how designers use the AI tools in practice. For instance, P4, a graphic designer whose primary work supports visual-related tasks, use AI primarily for generating different color combinations for the UI/UX design:

\begin{quote}
\textit{``
I work with AI most of the time because some color, I want to generate some color and some text.''}
\end{quote}

\subsubsection{Prompting Strategies for UI Design}

\para{Individual-level strategies}
%
%
%
Based on the interviews and participants’ demonstrations of using \generator{}s, we found that a variety of individual-level strategies are adopted to create prompts and use them when interacting with \generator{}s.
%
%
Some designers (P3, P6, P7, P9, P14, P16) create prompts using a \uline{template-driven strategy}, following predefined structures or patterns to guide their prompts with \generator{}s.
%
%
%
For instance, P14 consistently used a reusable template such as:
\textit{``Could you help design [task purpose]? Include [detail (such as what elements the UI will have)].''}
This template explicitly specified the task objective and the expected level of detail, allowing the designer to efficiently generate prompts by only modifying the task purpose while keeping the overall structure unchanged.
As P14 explained:
\begin{quote}
\textit{``I use it [\generator{}] for templates generation… I could also paste my templates to it, more like template cloning. It's more time efficient.''}
\end{quote}

Several others (P1, P12, P13, P15) adopted a \uline{context-driven strategy}, taking into account contextual factors such as visual themes, functional constraints, and stylistic requirements of the overall project. For example, P1 illustrated how these contextual factors directly influence their prompt decisions:

\begin{quote}    
    \textit{``Creating prompts is actually based on what you’re dealing with... For example, every application has different themes. If we’re working on a website with a green theme, the background may be green. Or if the site is mostly red, our warning will not be in red because it would blend into the background. So it really depends on the case at hand.''}
\end{quote}

Participants (P6, P9, P13, P16) also highlighted that creating effective prompts requires a \uline{human factor-driven strategy}, which involves clearly communicating the user needs, requirements and expectations of UIs through prompts. This is to ensure that the generated UIs are user-centered. For example, P6 emphasized the importance of making prompts explicitly centered on the needs of users:

\begin{quote}
    \textit{``I think all the considerations is mostly clarity, accessibility, and user control... making sure users can easily understand these options and manage preference without offering pressure or risk.''}   
\end{quote}

Besides what to include in prompts, participants also reported various strategies in terms of how to use prompts with \generator{}s. P10 actively \uline{adjusted the level of detail included in prompts} to enable UI design exploration. They not only crafted highly detailed prompts specifying functional and visual constraints, but also employed more general prompts to foster broader ideation and the generation of diverse design concepts. As P10 noted:

\begin{quote}
\textit{“Besides very specific prompts, I also give it [\generator{}] some very general ones as a Plan B, which let it freely improvise.”}
\end{quote}
By varying the level of detail in prompts, designers can flexibly switch between constraint-driven generation and open-ended exploration. Highly specific prompts help refine and validate concrete design requirements, while more general prompts serve as a “plan B,” allowing the generative system to freely explore alternative layouts or design directions beyond the designer’s initial assumptions.

%
In terms of using the prompts, some designers (P5, P10) \uline{repeated the same prompt across multiple generations} to explore possible design variations while accounting for the inherent uncertainty of \generator{} outputs.
For example, P10 described opening multiple \generator{} windows and submitting identical prompts in parallel to obtain different layouts or stylistic outcomes, such that he could compare the diverse generated results.
P10 noted:

\begin{quote}
    \textit{``Sometimes I open several windows and give it [\generator{}] the same prompt to let it generate multiple times. Usually I get different results.''}
\end{quote}
By producing multiple variations from an identical prompt, this approach offers designers a pool of alternatives that support comparison, informed selection, and the appropriation of appealing design elements and creative ideas.

In addition to text-based prompting, some participants (P2, P10) also mentioned that they supplement their prompts with visual materials when using \generator{}s, adopting what we call a \uline{multimodal prompting strategy} to combine text and visual inputs for richer design exploration.
For instance, P2 explained that designers may also incorporate images or diagrams when prompting, stating that 
\begin{quote}
\textit{``You can add some images from your laptop that you want to incorporate into the design you are creating when using Figma Make, and there should be an option to add images or diagrams.''}
\end{quote}
By incorporating visual inputs alongside text prompts, designers can more effectively convey structural and stylistic intentions, enabling the generative system to produce UI designs that are better aligned with reference designs and design goals.

\para{Assisted and collaborative strategies}
Beyond working independently, some participants relied on collaboration or used tool-assisted support, to guide prompt creation and ensure alignment with project goals.

In more formal or collaborative projects, some designers employed a \uline{co-creation prompting strategy} before submitting prompts to \generator{}s. Participants (P5, P10, P11) described the need to discuss design goals, task constraints, and functional requirements with project managers or software engineers to ensure that the prompt accurately reflected the project's intent and minimized the risk of misaligned outputs. In particular, P10 chose to use a \generator{} to demonstrate and generate an account deletion feature as a concrete example during the discussion. As P10 explained:

\begin{quote}
\textit{``If I were really creating such a feature [account deletion flow] in a work setting, I would first check with the program manager to confirm the context. What exactly it means to ‘remove their account from service’, and clarify those details before writing prompts.''}
\end{quote}

In addition to human collaborators, participants also treated AI agents as collaborators in the design process. They typically relied on two types of AI support: external general-purpose tools they regularly used, and the \generator{} itself. These AI collaborators supported iterative refinement, helped clarify requirements, and expanded design possibilities during prompt creation.
Specifically, several participants (P2, P10) adopted an \uline{multi-AI prompting strategy}, in which they first used general-purpose LLMs (e.g., ChatGPT) to structure, refine, or expand their prompts, and then transferred the prompts into Figma Make for UI generation. Participants reported that this practice helped clarify their intent, generate richer design flows, and selectively extract useful components before passing them to \generator{}s.
%
As P10 described:

\begin{quote}
\textit{``
I have used some GPTs for generating prompts...
%
I usually just let the normal GPT help me write. I will tell it my goal, and then let GPT help me write the prompt, and look at what it gives back. 
Usually it gives out a lot of content, and then from a designer’s perspective, I analyze whether it matches my expectations. Then I decide whether to revise the prompt or not, and if it is appropriate, I feed it into something like Figma Make. 
%
%
I work on B2B projects, I tend to let GPT generate the entire flow, from the first step to the next step. It will directly give me headlines, descriptions, and many details, like on-screen text. It also considers accessibility and emotion. I think this is quite good. Then I copy this part and give it to Figma Make to see what it could generate.''}
\end{quote}

%

In addition, some other participants (P5, P17) adopted a \uline{designer-AI collaborative strategy}, in which they iteratively guided and refined the textual design plans produced by \generator{}s before generating the final UI. Designers first stated the purpose and constraints of the task in plain language, then asked the \generator{}s to produce multiple textual design plans (e.g., step-by-step flows and layout suggestions) for comparison. They reviewed these plans, selected or combined preferred options, and repeated this select-and-refine cycle until satisfied with a chosen plan. Only after converging on a plan did they prompt the \generator{}s to produce concrete UI designs based on that plan.

\subsection{Designers' Attitude toward \generator{}s}

\subsubsection{Perceived Benefits of \generator{}s}

\para{\generator{}s help designers with brainstorming and provide inspiration}
Multiple participants (P1–P8, P10, P13, P16, P17) reported that \generator{}s are especially helpful during early-stage ideation, offering quick inspiration such as layout concepts, color palettes, text drafts, and multiple prototype variations. 
These tools help designers overcome creative barriers, such as feeling stuck or limited to familiar ideas, and support the exploration of diverse directions by quickly generating multiple concepts and alternatives that designers might not have envisioned on their own.
As P4 emphasized the breadth of ideas \generator{}s can offer:
\begin{quote}
\textit{``I use AI for my workflow and it has been a game changer for overcoming creative block.''}
\end{quote}

Some participants (P5, P10) particularly valued the ability of \generator{}s to generate diverse variations for comparison, evaluation, or user testing. 
As P5 stated:
\begin{quote}
\textit{``There could be many different variations of prototypes that can be generated by it [\generator{}] so we can show users and see if the workflow makes sense.''}
\end{quote}

\para{\generator{}s reduce human effort through automation and make design work easier}
Some participants (P1, P15–P17) highlighted that \generator{}s simplify design tasks by reducing repetitive activities, such as drafting layouts, generating text content, or iterating variations, and lowering cognitive load, making routine work less demanding.
Several participants appreciated how easily these tools integrate with their workflow and automate tedious steps. 
For example, P1 noted:

\begin{quote}
\textit{``With the assistance that AI provides, it really makes work easy. You don’t have to do everything yourself when designing frames, and there are many templates you can choose from. That is why I prefer AI-powered design tools.''}
\end{quote}



\para{\generator{}s help speed up the design process }
A recurring benefit mentioned by some participants (P6, P7–P9, P13–P15) is increased speed. \generator{}s accelerate both creative exploration and practical execution. For example, tasks that previously required several days to generate a UI can now be completed in 10–15 minutes by \generator{}s, enabling designers to deliver faster.
P9 noted this advantage:

\begin{quote}
\textit{``They speed up my whole process when it comes to developing and designing.''}
\end{quote}

\para{\generator{}s outputs are appealing}
Every participant expressed positive impressions of the UIs generated by \generator{}s. 
Although the UIs were not always production-ready, their visual aesthetics were often highly polished and met or exceeded expectations.
Participants consistently described the generated screens as clean, appealing, or surprisingly polished, demonstrating broad satisfaction with the visual appeal of \generator{}-created interfaces. As P15 noted:

\begin{quote}
    \textit{``The UI is okay for building the final interface. It is great, simple, and also very easy to understand.''}
\end{quote}

\subsubsection{Perceived Drawbacks of \generator{}s}
%

\para{\generator{}s often provide generic and repetitive output, causing off-brand UI designs}
%
%
Several participants (P6–P8, P10, P15) noted that \generator{}s frequently produce designs that look generic, repetitive, which are insufficiently aligned with a product’s brand identity. 
Specifically, these outputs often reuse similar layout structures, color schemes, typography styles, and component arrangements across different designs, resulting in interfaces that look visually similar regardless of the product context. Rather than reflecting brand-specific characteristics such as unique visual tone, personality, or target user group, the generated designs tend to resemble standard templates or “default” app styles.
These outputs tended to follow repetitive patterns or templates, which might stem from common design conventions in the datasets~\cite{uizard_research, figma_ai_blog} used to train \generator{}s. This limitation is not due to a lack of visual appeal, but rather because the resulting UIs are overly generic and insufficiently distinctive for projects that require a strong, differentiated visual direction.
As P15 explained:

\begin{quote}
\textit{``What I dislike is that it doesn’t usually understand complex design and textures or context. Some tools limit creativity because they follow repetitive patterns or templates.''}
\end{quote}

\para{\generator{}s often generate UIs with incoherent color choices}
Participants (P14, P16) also reported that \generator{}s sometimes generate layouts with inconsistent structures or poorly chosen color palettes that lack aesthetic coherence and a clear sense of visual style. 
Specifically, the generated colors often do not work harmoniously together, resulting in interfaces that feel visually unbalanced, stylistically inconsistent, or lacking an overall “design touch.”
These issues suggest that \generator{}s lack awareness of visual hierarchy and the nuanced relationships among colors in UI design. 
As a result, designers must invest substantial manual effort to correct the generated outputs, such as adjusting colors, contrast, and alignment to achieve a cohesive and professional appearance.
As P16 described:

\begin{quote}
\textit{``I could say it doesn’t really have this clear sense of touching when it comes to coloring, like the appearance of websites in terms of the color palette. So most of the time I have to be the one correcting the colors.''}
\end{quote}

\para{Limitations of \generator{}s in prompt understanding}
Participants (P9, P10, P11, P14, P15) reported that \generator{}s often fail to fully understand their prompts, such as the notions about user experience. P11 noted that several rounds of prompting are typically required before obtaining an acceptable structure, as the generated results often lack UX focus:

\begin{quote}
\textit{``I need many rounds of prompts to get the structure I want... the structure it gives is often not UX-focused.''}
\end{quote}

This misalignment suggests that \generator{}s may struggle to accurately interpret high-level design goals and translate them into appropriate interface structures. As a result, participants frequently needed to issue repeated correction prompts to fix the generated outputs. P9 described this as a continuous cycle of correction: when the generated UIs deviated from their intended design, he should repeatedly prompt the generator to adjust the outputs, which he found exhausting.
%

However, P10 noted that for complex design problems, relying on iterative prompting was often ineffective, as repeated prompt adjustments rarely yielded the intended outputs. In these situations, participants found it more efficient to manually edit the designs, especially since the \generator{}s' outputs could be exported directly into other design tools such as Figma.
%
%
P14 additionally observed that even with carefully crafted prompts, \generator{}s may still make unexpected or inexplicable mistakes, indicating limitations in their ability to fully digest user instructions:

\begin{quote}
\textit{``Well, most of the problem I encountered is about the AI not fully following my prompts. You know, fully digesting what I mean in my prompt. Because there are times when, even if you try as hard as you can to make your prompt perfect, it still ends up making slight mistakes, or even significant mistakes, you don’t know why, that you probably need to edit or correct on your own.''}
\end{quote}

Together, these findings indicate that limitations in prompt understanding and design intent interpretation significantly hinder the effectiveness of \generator{}s in end-to-end UI design. To better support practical workflows, \generator{}s should not only provide more robust and controllable correction mechanisms, but also offer greater transparency in their generation processes (e.g., exposing intermediate decisions or layout rationales) to help users understand, diagnose, and more effectively guide the system.

\para{UI designers find it difficult to reuse and maintain generated UI code}
Unlike traditional UI design, where designers and developers work separately, some \generator{}s can generate code (such as HTML~\cite{html_intro}, CSS~\cite{css_intro}, or React components~\cite{react}) directly alongside the UI design during the development process.
However, the generated code often has several practical issues. For example, Figma Make can export UI designs as HTML/CSS or React code, and it can even handle mobile layouts. However, the generated code is intended for web platforms, not native mobile applications. Mobile platforms like Android and iOS use Java~\cite{java}/Kotlin~\cite{kotlin} or Swift~\cite{swift}/Objective-C~\cite{objectivec}, so the generated HTML/CSS/JavaScript~\cite{javascript} code cannot be directly reused in native apps. Even when generating responsive mobile layouts, developers must rewrite most of the code to fit the mobile platform’s requirements, including touch interactions, screen scaling, and performance optimization. As a result, developers usually cannot use the exported code directly and must spend significant effort refactoring it for maintainability, responsiveness, and production readiness.
Participants (P5, P10) raised concerns about the maintainability of such AI-generated code, noting that extensive cleanup is usually required before it can be integrated into production. Consequently, the practical applicability of these generators in production-level UI design is limited.
As P5 described:
\begin{quote}
\textit{``It generates code that is still hard to maintain... you have to chase through all these things, and that’s very tedious.''}
\end{quote}

%

\para{Human intervention is essential in \generator{}s-Generated UI designs}
%
%
\generator{}s have several drawbacks, which prevent designers from relying on them entirely. Some participants (P2, P4, P8, P9, P10) mentioned that they treat AI-generated UIs as initial drafts rather than final solutions. 
They generally see AI as a supportive tool that can provide inspiration, accelerate ideation, or generate alternative options, but not as a complete replacement for their own decision-making. 
For example, P9 noted:

\begin{quote}
    \textit{``Of course, I don’t want to over-rely on AI, because in my line of work, once you are using AI, you tend to forget some things. So I’m trying to just use them to sharpen my ideas, not to fully rely on them.''} 
\end{quote}

This limited trust in \generator{}s is closely related to the perceived need for human judgment and intervention. Participants (P6, P8, P13, P15) consistently emphasized that AI-generated outputs rarely reach production quality on their own and require substantial refinement by designers. 
They reported the need to correct errors, adjust details, and incorporate domain knowledge and contextual understanding that AI systems often lack. Designers stressed the importance of applying their personal judgment, professional expertise, and creative vision to ensure that the final design is not only technically correct but also aligns with project goals, user needs, and situational constraints. In this sense, \generator{}s are positioned as assistive tools rather than autonomous designers, with human decision-making remaining central throughout the design process. P8 explained:

\begin{quote}
    \textit{``Even though I see AI as a supporting tool, I believe in myself as the main driver of my design. I just use it when I try to get inspiration about what I should probably do. I usually prefer to rely on my own judgment and creativity to refine things.''} 
\end{quote}

\ignore{\para{AI-generated outputs may pose privacy risks}
Beyond functional limitations, participant (P17) expressed concerns about the privacy and security of outputs generated by \generator{}s. 
P17 noted:
\begin{quote}
\textit{``What I dislike is that AI tools sometimes lack privacy… they cannot really protect the user. For instance, AI can generate images that are not real but look very realistic.''}
\end{quote}
}


\para{Concerns about personal data privacy and security when using \generator{}s}  
Several participants (P7, P9–P11) expressed concerns about the privacy and security of sensitive or proprietary design data when using \generator{}s. They noted that uploading client-related content or internal project materials to AI tools could risk exposure or misuse. 
P11 emphasized the confidentiality of the data shared with AI tools:

\begin{quote}
\textit{``I kind of worry about the security of the kind of confidentiality of the stops I share with my AI because of privacy reasons.''}
\end{quote}

P10 also described limitations with enterprise projects and the inability to freely upload internal design references due to confidentiality:
\begin{quote}
\textit{``Many of our internal enterprise projects are complex. Due to confidentiality, previous designs cannot be directly used with AI tools.''}
\end{quote}

\para{Overreliance on \generator{}s can reduce personal creativity}  
While participants (P8, P12) acknowledged that \generator{}s are valuable for supporting brainstorming, they also expressed concerns about the risks of over-reliance on AI, particularly when it replaces rather than augments their own creative thinking. Specifically, participants worried that excessive dependence on AI-generated outputs could reduce their active engagement in the design process and weaken the development of personal style and design reasoning.
P12 highlighted that relying too heavily on AI may impede designers from fully understanding how to design, emphasizing the importance of first forming their own ideas before turning to AI for inspiration:
\begin{quote}
\textit{``If you rely too much on AI usage, there's a tendency that you might not fully know how to design properly. So I try to build on myself first, then use it to generate ideas.''}
\end{quote}

\subsection{Organizational Climate toward \generator{}s}

\para{Use of enterprise-hosted AI to protect intellectual property}
Some participants (P9, P10, P11), all of whom reported working in large companies, mentioned that their organizations have AI security policies in place to prevent the exposure of sensitive enterprise data.
Their companies explicitly prohibit uploading internal data to unapproved AI tools, and only AI systems that have passed internal security reviews are allowed to access and analyze company data. 
They explained that these measures are intended to prevent confidential company information from being used to train external AI models or being exposed to third-party platforms. 
For example, P11 explained that their company built an internal enterprise AI system to replace the use of external AI tools and to ensure that all data remains within the company’s secure infrastructure:

\begin{quote}
\textit{``We are not allowed to use external GPT tools, because putting any company materials into external tools for training is considered illegal. So the company built its own enterprise AI system. The data is protected inside the enterprise, and the company uses its own internal materials for training. You are not allowed to copy production files into external software, because those tools have not passed the company’s network security review. We can use AI for idea exploration, but we still cannot directly put production files into it for modification.''}
\end{quote}

Similarly, P10 described how their company mandates the use of enterprise versions of AI tools and strictly restricts external platforms; if they want to use an unauthorized tool, they must obtain approval from their company's IT department.

\begin{quote}
\textit{``If you want to introduce a new AI tool, you have to report it to IT for approval. The IT department will review the permissions and security policies, and if it looks suspicious, they will not allow it to be used with company data.''}
\end{quote}

\para{Requiring human oversight of \generator{} usage to preserve design control}
Some participants from small teams (P2, P6, P12, P17) reported that their organizations generally encourage the use of AI tools to support design work, but explicitly emphasize the importance of human oversight and caution against fully relying on AI. In these teams, AI is viewed as a helpful assistant for improving efficiency and speeding up workflows, rather than a substitute for designers’ expertise and judgment. For example, P2 stated:

\begin{quote}
\textit{``Yes, my company supports the use of AI tools together with your knowledge, not 100\% AI though.''}
\end{quote}

Similarly, P6 explained that while their team encourages the use of AI to enhance productivity, clear boundaries are maintained to ensure that human judgment remains central to the design process:

\begin{quote}
\textit{``Most of my team do encourage using AI tools to speed workflow and everything, but we have clear limits on it. AI can actually assist, but not replace human judgment and not take control of the whole work. We always review and refine AI outputs to ensure accuracy and functionality.''}
\end{quote}

\para{Unregulated AI Practices}
In contrast, some participants from other small teams (P1, P3) reported that their organizations have no formal policies or restrictions regarding the use of AI tools. As P1 noted, there is currently no existing policy governing AI usage in their team, indicating a more open and less regulated approach to AI adoption in certain small-team contexts.

\section{Discussion}
\label{sec:discussion}

\subsection{Suggestions Toward Better \generator{}s for UI Designers}
%
%
Our study reveals that \generator{}s reshape UI design workflows by enabling designers to accelerate ideation and prototyping through a new form of human–AI collaboration.
At the same time, the study identifies difficulties and unexpected efforts that designers encounter when using \generator{} in real-world settings (e.g., commercial products), suggesting opportunities to improve \generator{}s.


\para{\generator{} built-in prompting strategies to reduce the burden of prompt creation for designers}
%
Our findings show that designers adopt a wide range of prompting strategies when working with \generator{}s, both individually (e.g., template-driven or human factor-driven approaches) and collaboratively (e.g., co-creation prompting with team members). This diversity suggests there is no single optimal method for using \generator{}s; instead, designers rely on different approaches that require expertise and additional effort, especially for those newer to AI tools.
To help designers more effectively explore and select prompts that serve their UI design tasks, we suggest that \generator{}s provide built-in prompting strategies (e.g., reusable prompt templates) that designers can freely adopt. 
Such features could lower the entry barrier for less experienced designers and reduce the trial-and-error effort typically required to find effective prompts. 
Additionally, built-in strategies can encourage more consistent and efficient human–AI collaboration across different design contexts.

\para{Enabling designers to explicitly attribute resulting UI designs to different parts of the prompts}
%
%
Our findings show that \generator{}s may misinterpret designers' prompts, and even carefully crafted prompts can lead to erroneous or unexpected results, forcing designers into time-consuming trial-and-error. This effort is further complicated by the lack of transparency in \generator{}s regarding which parts of a prompt are likely to cause problems in the generated UI, i.e., attribution.
Therefore, one of our suggestions for \generator{}s is to improve transparency in the UI design generation process by explicitly indicating how different parts of the prompt contributed to elements in the generated UI, potentially by adopting state-of-the-art attribution techniques~\cite{wang2025tracllm,cohen2024contextcite}.
By making these prompt-output relationships explicit, such features would help designers more easily identify which aspects of a prompt require revision, thereby streamlining iteration and reducing unnecessary experimentation.

%

\para{Shifting from generic UI generation to designs that align with specific brand profiles}
In practice, designers expect \generator{}s to produce UI designs that reflect the brand identity and visual preferences of their companies.
However, although \generator{}s support multiple input modalities (e.g., textual descriptions and reference images), they do not inherently allow designers to communicate brand requirements effectively and generate UIs that meet those requirements.
To address this issue, we recommend that \generator{}s shift from generating generic, sometimes repetitive UI designs to producing specific designs that align with each company’s requirements.
For example, \generator{}s may allow designers to define custom brand profiles that specify color palettes, typography, layout conventions, and component styles, and then apply these profiles consistently across generations. 
Making such constraints explicit and reusable would not only reduce off-brand outputs but also improve designers’ confidence and trust in \generator{}s' ability to support professional design workflows.

\para{Enabling reuse of \generator{}-generated UI code via modular and well-documented outputs}
Participants also raised concerns about the difficulty of reusing and maintaining UI code generated by \generator{}s. 
Generated outputs were often perceived as poorly structured or insufficiently documented, requiring substantial manual refactoring before they could be incorporated into real-world projects. 
These challenges reduced the practical value of generated designs, particularly in collaborative and long-term development contexts.
To improve reusability, future \generator{}s should prioritize generating modular, well-structured UI code that aligns with common design and development practices. 
Producing component-based code, following established naming conventions, and providing clear documentation would allow designers to more easily adapt and extend generated outputs. 
Improving code quality in this way would help position \generator{}s as viable tools not only for rapid prototyping, but also for integration into production workflows.

%

\para{Adding mechanisms to better inform designers about the data practices of \generator{}s}
Concerns about data privacy and security further influenced participants’ willingness to adopt \generator{}s, especially in organizational and enterprise settings. Designers were often uncertain about how prompts, design assets, or proprietary information were processed, stored, or reused by the system. This lack of transparency limited trust and discouraged the use of \generator{}s for sensitive or client-facing projects.
To mitigate these concerns, future \generator{}s should provide clearer communication and stronger safeguards around data handling practices. Possible approaches include offering local or private deployment options, enabling users to control data retention policies, and explicitly disclosing how design data is used within the system through accessible and user-friendly interfaces. For example, tools could present clear consent dialogs or onboarding notices, similar to website cookie consent banners or mobile privacy disclosures that explain, in plain language, what data is collected, how it is processed, and whether it is retained or shared. Strengthening privacy protections and improving the accessibility of privacy-related information would increase designers’ confidence and support broader adoption of \generator{}s in professional environments.

\para{Establishing organizational policies that balance open adoption and responsible use of \generator{}s}
An interesting observation is that, at the organizational level, attitudes toward \generator{} adoption diverge across organizations.
For example, some emphasize the importance of enterprise-hosted AI and human oversight, while others embrace open, unregulated AI practices.
These choices have different implications for both the safe use of AI and work efficiency, although participants did not discuss them systematically.
As \generator{} use becomes more common among designers, we believe that formal guidelines, policies, or governance mechanisms are essential to balance open adoption (for efficiency) and responsible use (for safety).
In particular, these resources would greatly benefit small- and medium-sized organizations. 
Even without institutional AI experts, such organizations could use these resources to make informed decisions about which \generator{}s to use for different UI design tasks, how to use them responsibly, and how to avoid potential risks while maintaining work efficiency.

%

\para{Balancing AI assistance with the preservation of designers' professional skills}
 While \generator{}s offer clear benefits in accelerating ideation and reducing repetitive effort, our findings also surface a subtler long-term concern: the risk that sustained reliance on AI-generated outputs may gradually erode designers' own creative and technical skills. Participants explicitly flagged this tension, noting that over-reliance on \generator{}s could lead designers, particularly those early in their careers, to bypass the effortful practice through which design judgment is typically developed. As P12 noted, heavy AI use risks producing designers who struggle to design independently, undermining the deep craft knowledge that underlies professional expertise. 
This connects to broader HCI and learning-science debates about how tool-mediated work shapes skill acquisition: when AI handles the generative and exploratory phases of design, designers may have fewer opportunities to engage in the productive struggle that cultivates intuition, visual reasoning, and contextual judgment -- capabilities essential for both high-quality creative work and effective use of AI tools.
We therefore suggest that \generator{}s should be designed not only to assist designers in the moment, but also to support their longer-term skill growth. For example, tools might offer optional "ideation exploration" modes that encourage designers to articulate their own design rationale before revealing AI-generated alternatives, or provide post-generation reflection prompts that invite designers to critically compare AI outputs against their own initial thinking. Such features could help preserve the cognitive engagement that is central to developing design expertise, while still capturing the efficiency benefits that make \generator{}s valuable in professional practice.

\subsection{Limitations and Future Work}
This study explores how UI designers use \generator{}s in their design practices. 
One limitation of this work is that our findings are derived from a relatively small sample of participants, which may limit the generalizability of the results to broader design contexts.
We determined the sample size based on thematic saturation to reflect common practices reported by designers in real-world settings.
As such, our findings should be understood as a snapshot of the current landscape of designers’ interactions with \generator{}s, rather than a comprehensive account. Nevertheless, we believe the study captures several dominant design scenarios and sheds light on key challenges and opportunities faced by designers when integrating these tools into their workflows.
To address the above limitations, future work will extend this research by engaging with the developers of \generator{}s to better understand the design rationales, implementation choices, and underlying assumptions embedded in these systems. Such investigations could help bridge the gap between tool design and designers’ needs, and inform the development of more effective and responsible generative design tools.

\section{Conclusion}
\label{sec:conclusion}

This study investigates how prompt-to-design generators (\generator{}s) are used and perceived in everyday UI design practice, with particular attention to their perceived benefits and concerns. Through interviews with 17 professional UI designers, we found that designers primarily use \generator{}s for ideation and prototyping, selectively incorporating outputs into established practices rather than relying on them for end-to-end design. Designers appreciate the efficiency and inspiration these tools provide but remain concerned about potential losses in personal creativity and design quality. We further find that organizational climate also influences the adoption and governance of \generator{}s. Our work highlights the need for improved tool capabilities and organizational policies and guidelines that support sustainable, responsible, and designer-centered use of GenAI tools.

\bibliographystyle{ACM-Reference-Format}
\bibliography{ref}

\appendix
\section{Additional Tables and Figures}
\label{sec:appendix_a}

\begin{table*}[t]
\centering
\small
\caption{Participants’ Demographic Information}
\label{tab:participant_info}
\begin{tabular}{ccccc}
\toprule
\textbf{No.} & \textbf{Company Scale} & \textbf{Age} & \textbf{Years of UI/UX Design} & \textbf{Commonly Used \generator{}s} \\ 
\midrule
\rowcolor[gray]{.8}
P1 & Small Team & 18--25 & 3--5 & Stitch, Uizard \\ \hline
P2 & Small Team & 26--35 & 3--5 & Figma Make \\ \hline
\rowcolor[gray]{.8}
P3 & Small Team & 18--25 & 3--5 & Magician (Figma plugin) \\ \hline
P4 & Small Team & 26--35 & 5+ & Builder.io, Framer, Uizard \\ \hline
\rowcolor[gray]{.8}
P5 & Large Company & 46--60 & 5+ & Bolt, Lovable \\ \hline
P6 & Small Team & 18--25 & 3--5 & Figma Make, Uizard \\ \hline
\rowcolor[gray]{.8}
P7 & Freelancer & 18--25 & 3--5 & Uizard \\ \hline
P8 & Freelancer & 26--35 & 3--5 & Magician (Figma plugin), V0 \\ \hline
\rowcolor[gray]{.8}
P9 & Large Company & 18--25 & 1--2 & Figma Make, Lovable \\ \hline
P10 & Large Company & 26--35 & 5+ & Figma Make, Lovable, Uxpilot, V0 \\ \hline
\rowcolor[gray]{.8}
P11 & Large Company & 26--35 & 3--5 & Figma Make, Lovable \\ \hline
P12 & Small Team & 18--25 & 1--2 & Adobe Firefly \\ \hline
\rowcolor[gray]{.8}
P13 & Large Company & 18--25 & 1--2 & Adobe Firefly, Framer \\ \hline
P14 & Large Company & 18--25 & 5+ & Figma Make, Framer \\ \hline
\rowcolor[gray]{.8}
P15 & Small Team & 26--35 & 3--5 & Figma Make, Lovable \\ \hline
P16 & Freelancer & 26--35 & 5+ & Uizard, V0 \\ \hline
\rowcolor[gray]{.8}
P17 & Small Team & 36--45 & 5+ & V0 \\
\bottomrule
\end{tabular}
\end{table*}

\begin{table*}[t]
\caption{Interview Questions List}
\label{tab:interview_list}
\centering
\small
\begin{tabular}{p{3cm} p{13cm}} \\
\hline
\textbf{Warm-up Interview} \newline \textbf{Questions}&
1. How many years of experience do you have in UI/UX design? \newline
2. What is the size of your company or team? What is your role? \newline
3. What types of interfaces do you primarily design (e.g., web, mobile apps, or multiple platforms)? \newline
4. Which non-AI UI design tools do you usually work with? \newline
5. What AI-powered design tools (\generator{}s) do you use when designing UIs? \newline
\hspace*{1em} a. How often do you use them? \newline
\hspace*{1em} b. Which ones do you use most often? \newline
\hspace*{1em} c. Why do you use them? \newline
\hspace*{1em} d. Roughly what percentage of your design process involves AI-generated interfaces? \newline
\hspace*{1em} e. In what ways do you typically use AI to support your work? \newline
\hspace*{1em} f. What do you like about \generator{}s? \newline
\hspace*{1em} g. What do you dislike about \generator{}s? \newline
\hspace*{1em} h. What problems do you encounter when using \generator{}s? \\[1em]
\hline

\textbf{Follow-up Interview} \newline \textbf{Questions} &
1. How do you feel about the UI that was generated by \generator{}s? \newline
2. Would the UI generated by \generator{}s be ready for direct use in a real-world application? Why or why not? \newline
%
3. What considerations did you have in mind when designing the prompts? \newline
4. What key information do you think should be included in the prompts? \newline
5. Which part of the prompt did you find most important or spend the most time crafting? \newline
6. Did you modify the prompts based on the AI-generated results? If yes, how and why? \newline
7. What challenges did you face when designing the prompts? \newline
\hspace*{1em} a. If you faced challenges, what kind of support or guidance would have helped? \newline
%
8. Besides the prompt, did you use any other settings or features of the \generator{}s? \newline
9. What is your company’s policy regarding the use of \generator{}s? \newline
10. Has your company purchased an enterprise version of any \generator{}s? \\
\hline

\end{tabular}
\end{table*}

\end{document}